\documentclass[preprint,pre,showpacs,amsmath,amssymb,superscriptaddress,aps]{revtex4-1}

\usepackage{amsmath,amssymb,graphicx,dcolumn,bm,hyperref,verbatim,color}
\usepackage{longtable,subfigure,wrapfig,epsfig,float,array,psfrag,bbold}

\definecolor{azure}{rgb}{0.0, 0.5, 1.0}
\definecolor{asparagus}{rgb}{0.53, 0.66, 0.42}
\definecolor{ballblue}{rgb}{0.13, 0.67, 0.8}
\definecolor{sgreen}{rgb}{0.0, 0.8, 0.35}
\definecolor{darkgreen}{rgb}{0.0, 0.5, 0.0}
\definecolor{sred}{rgb}{0.9, 0.6, 0.4}
\definecolor{purp}{rgb}{0.4,0.1,0.7}
\definecolor{webgreen}{rgb}{0,.35,0}
\definecolor{webbrown}{rgb}{.6,0,0}
\definecolor{royalblue}{rgb}{0,0,0.9}

\usepackage{tikz}
\renewcommand{\vec}[1]{\mathbf{#1}}

\newcommand{\p}{\partial}
\newcommand{\mcA}{{A}}
\newcommand{\mcF}{\mathcal{F}}

\hypersetup{
   colorlinks=true, linktocpage=true, pdfstartpage=3, pdfstartview=FitV,
   breaklinks=true, pdfpagemode=UseNone, pageanchor=true, pdfpagemode=UseOutlines,
   plainpages=false, bookmarksnumbered, bookmarksopen=true, bookmarksopenlevel=1,
   hypertexnames=true, pdfhighlight=/O,
   urlcolor=webbrown, linkcolor=royalblue, citecolor=webgreen,
   pdfauthor={Christoph Weber, Chris H. Rycroft, and L. Mahadevan},
   pdfsubject={},
   pdftitle={Differential Activity drives Instabilities in Biphasic Active Matter},
   pdfkeywords={instabilities in active systems, hydrodynamic instabilities, multi-phase theories,
 contractions of active gels},
   pdfcreator={pdfLaTeX},
   pdfproducer={LaTeX with hyperref}
}

\begin{document}

\title{Differential-activity driven instabilities in biphasic active matter}

\author{Christoph A.\ Weber}
\affiliation{Paulson School of Engineering and Applied Sciences, Harvard
University, Cambridge, MA 02138, USA}
\author{Chris H.\ Rycroft}
\affiliation{Paulson School of Engineering and Applied Sciences, Harvard
University, Cambridge, MA 02138, USA}
\affiliation{Mathematics Group, Lawrence Berkeley National Laboratory,
Berkeley, CA 94720, USA}
\author{L.\ Mahadevan}
\affiliation{Paulson School of Engineering and Applied Sciences, Department of Physics, Department of Organismic and Evolutionary Biology, Harvard
University, Cambridge, MA 02138, USA}

\begin{abstract}
Active stresses can cause instabilities in contractile gels and living tissues. Here we describe a generic hydrodynamic theory that treats these systems as a
mixture of two phases of varying activity and different mechanical properties. We find that differential activity between the phases provides a mechanism causing a demixing instability.
We follow the nonlinear evolution of the instability and characterize a phase diagram of the resulting patterns. 
Our study 
complements other instability mechanisms in mixtures 
such as  differential growth, shape, motion or adhesion. 
\end{abstract}



\maketitle

Biological systems are distinguished by the presence of active stresses which can affect their physical properties and alter their stability. For example, active stresses  give rise to collectively moving streaks~\cite{Schaller} and clusters~\cite{Dombrowski_2004, pnas_bacteria,VibratedDisks2013},  rotating ring, swirl or aster-like patterns~\cite{Backouche_phase_diagram_active_gels,Kudrolli_2008,Yutaka,Ringe_PNAS, mori2011intracellular, Needleman_microtubule_contraction}, or the remodelling of cell-to-cell junctions in living tissues~\cite{rauzi2010planar}. These systems are typically described as a single phase with active stresses that drive the assembly of the constituents and the properties of the phases are typically assumed as liquid-like~\cite{bois2011pattern,Toner_2012, prost2015active} or even gases~\cite{Aronson_MT, Bertin_long, Weber_NJP_2013, thuroff2013critical}. 

\begin{figure}[tb]
  \centering
  \includegraphics[width=0.8\textwidth]{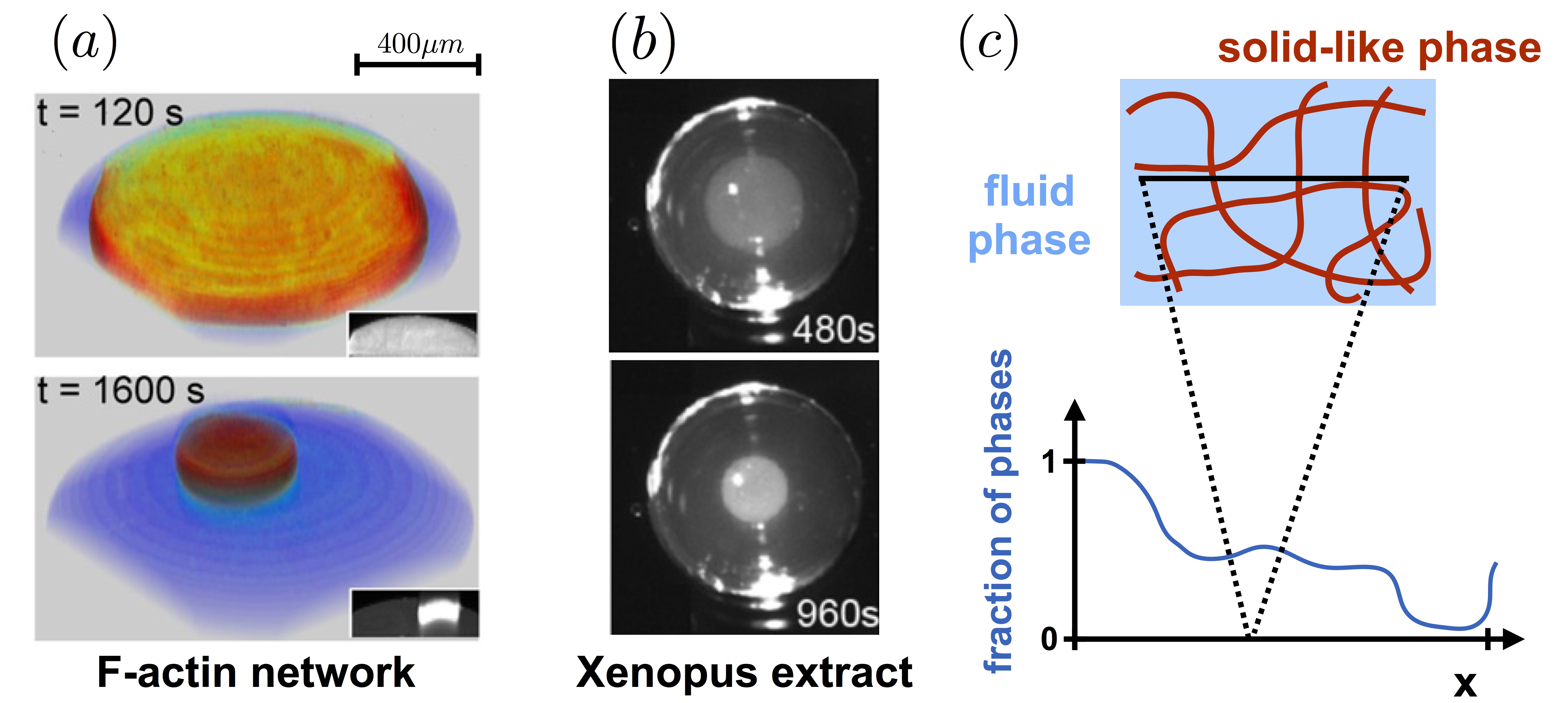}
  \caption{\label{fig:Fig_1}Examples of unstable, biphasic mixtures and
  illustration of two-phase mixture theory. (a) Contractile behavior of
  actin-myosin II network crosslinked with $\alpha$-actinin, and (b)
  contracting Xenopus egg extract; both are taken from
  Ref.~\cite{Bendix20083126}. (c) Illustration of biphasic mixture theory.
  Microscopic illustration of a biphasic gel (top), and hydrodynamic
  representation on large wavelength described by the volume fraction field of
  the constituent phases (bottom). The theory is valid on length scales above
  the pore size. On these scales, hydrodynamic variables are continuous, impact of microscopic interfaces is captured as friction proportional to the velocity difference~\cite{skotheim2004dynamics}, and the isotropic part of
  the activity should dominate any anisotropic contribution since activity is typically generated on or below the scale of the pore.}
\end{figure}

However many active material cannot be treated as fluids. Examples include cartilage, bone, tissues in early development~\cite{Barna2007931,Tabler20164,Martinjcb.200910099}, and superprecipitated systems such as networks of filaments connected by crosslinks and molecular motors~\cite{mori2011intracellular, Bendix20083126, Needleman_microtubule_contraction}.  The presence of activity in these systems can drive the macroscopic contraction of gels (\cite{mori2011intracellular, Bendix20083126,Needleman_microtubule_contraction}, and Fig.~\ref{fig:Fig_1}(a,b)), the compaction of cells during the condensation of cartilage cells~\cite{Barna2007931}, the network formation of osteoblasts during skull closure in embryos~\cite{Tabler20164}, and the formation of furrows in tissues~\cite{Martinjcb.200910099}. This requires the augmentation of previous passive biphasic descriptions, such as associated with poroelasticity~\cite{maha_cytoplasm_poroelastic_material} to account for active stress regulation and diffusion in cells and tissues. While recent work has included activity in a poroelastic description~\cite{radszuweit2013intracellular,radszuweit2014active}, the material was assumed to be homogeneous and stable despite  active stress generation in one of the phases. Here we question this assumption of stability  of an active mixture composed of two phases with different mechanical properties, and ask under what physical conditions an active poroelastic material might contract/condense or disintegrate/fragment, a phenomenon seen in a variety of experimental systems (Fig.~\ref{fig:Fig_1}(a-c)).

We find that differential activity between the solid and fluid phases that constitute an initially homogeneneous poroelastic medium can drive a mechanical instability leading to the demixing of a homogeneous medium into condensed solid-like patches that arise due to macroscopic contraction, reminiscent of observations in superprecipitated gels and compacted cells in tissues.  Depending on the rate and ability of transport of material and stress in the biphasic material, we find both uniformly growing and pulsatile instabilities leading to assembling, disassembling and drifting solid-like clusters that undergo fusion and fission.

We start with a consideration of isotropic active systems composed of two immiscible phases, $i=1,2$. These systems can be described by a hydrodynamic theory similar to descriptions used for fluid-like biphasic matter~\cite{drew1983mathematical, keener2011kinetics}, or elastic~\cite{doi_review_swelling_and_volume_PT} and viscoelastic gels~\cite{tanaka1993unusual, tanaka2000viscoelastic}. This theory is valid on length scales above  the characteristic pore size of the solid-like phase (Fig.~\ref{fig:Fig_1}(c)).  At the simplest level, activity in our biphasic system is described as an isotropic active stress  that acts on each phase which responds to this stress  according to its passive mechanical properties which are either fluid or solid-like,
respectively. Each phase $(i)$ is described by the hydrodynamic variables of velocity $v^{(i)}_\alpha$, volume fraction $\phi^{(i)}$ and displacement
$u^{(i)}_\alpha$ with $\p_t u^{(i)}_\alpha = v^{(i)}_\alpha$. The overall system is assumed to be incompressible and fully occupied by the two phases, i.e.\ $\phi\equiv\phi^{(1)} =1- \phi^{(2)}$. The fractions of each phase $i$ are conserved, so that $\p_t \phi^{(i)} = - \p_\alpha (\phi^{(i)} v^{(i)}_\alpha + j^{(i)}_\alpha)$, where $j^{(i)}_\alpha$ denotes a relative flux between the phases with $j^{(1)}_\alpha= -j^{(2)}_\alpha =: j_\alpha$. This relative flux can for example stem from rare unbinding events of components that belong to one of the phases. The resulting unbound components can diffuse and thereby cause an effective diffusive flux of the bound components (see Supplemental Material~\cite{SM}, I). For simplicity, we write $j_\alpha=-D \p_\alpha \phi$, where $D$ denotes the diffusion constant. The two conservation laws can be equivalently expressed by one transport equation and an incompressibility
condition,
\begin{subequations}
  \begin{align}
    \p_t \phi &= - \p_\alpha \left( \phi v^{(1)}_\alpha \right) +D \p_\alpha^2 \phi \, , \\
    \label{eq:incompressibility_cond}
    0&= \p_\alpha \left[ \phi \, v^{(1)}_\alpha + \left( 1-\phi \right) v^{(2)}_\alpha \right] \, .
  \end{align}
\end{subequations}
Neglecting osmotic effects and inertia, force balance in each phase implies:
\begin{subequations}\label{eq_first_set}
  \begin{align}\label{eq:force_balance_boundary1}
    0&=\p_\beta \left( \phi \sigma_{\alpha \beta}^{(1)} \right) - \phi \p_\alpha p - \mcF_\alpha \, , \\ \label{eq:force_balance_boundary2}
    0&=\p_\beta \left( (1-\phi) \sigma_{\alpha \beta}^{(2)} \right) - (1-\phi) \p_\alpha p + \mcF_\alpha \, ,
  \end{align}
where $\sigma_{\alpha \beta}^{(i)}$ is the additional stress (beyond the pressure) in each phase, and we have assumed, as in mixture theory~\cite{barry2001asymptotic, cowin2012mixture,  skotheim2004dynamics} that this stress is weighted by the respective volume fraction~\cite{footnote2}. The pressure $p$ acts as a Lagrange multiplier that ensures the incompressibility condition Eq.~\eqref{eq:incompressibility_cond}.  Momentum transfer between the phases is described by a friction force density $\mcF$. To leading order $\mcF$ is proportional to the relative velocity of the phases, \smash{$\mcF_\alpha = \Gamma(\phi) ( v^{(1)}_\alpha - v^{(2)}_\alpha)$}, where $\Gamma(\phi)=\Gamma_0 \phi(1-\phi)$ is the friction coefficient between the phases with $\Gamma_0$ constant. The dependence of the friction coefficient on volume fraction is a consequence of the condition that hydrodynamic momentum transfer vanishes if one of the phases is absent, i.e.\ $\mcF=0$ for $\phi=0$ or $\phi=1$.  Finally, we additively decompose  the stress into the passive stress \smash{$\sigma_{\alpha \beta}^{(i),\text{p}}$} and the isotropic activity $\mcA^{(i)}$,
\begin{align}
  \sigma_{\alpha \beta}^{(i)} &= \sigma_{\alpha \beta}^{(i),\text{p}} + \mcA^{(i)} \, \delta_{\alpha \beta} \, ,
\end{align}
\end{subequations}
where the passive stress \smash{$\sigma_{\alpha \beta}^{(i),\text{p}}$}  characterizes the mechanical properties of each phase. In general, the activity depends on all hydrodynamic variables. For simplicity, we focus on activities that depend on the volume fraction $\phi$, $\mcA^{(i)} =\mcA^{(i)}(\phi)$. Eqs.~\eqref{eq_first_set}  can be rewritten as
\begin{subequations}\label{eq:all_equations}
\begin{align}\label{eq:force_balance_boundary1_2}
  0&=\p_\beta \left( \phi \sigma_{\alpha \beta}^{(1)} + (1-\phi) \sigma_{\alpha \beta}^{(2)}  - \delta_{\alpha \beta} p\right) \, , \\
  \nonumber
    0&=\p_\beta \left( \phi \, \sigma^{(1),\text{p}}_{\alpha \beta} \right)
  - \frac{\phi}{1-\phi} \p_\beta \left( (1-\phi) \sigma^{(2),\text{p}}_{\alpha \beta} \right)
  \\
  &+ \left( v^{(2)}_\alpha - v^{(1)}_\alpha \right) \Gamma_0 \phi
  +\phi \, {\mcA}(\phi) \p_\alpha \phi \, ,
  \label{eq:force_balance_boundary2_2}
\end{align}
where we define
\begin{equation}\label{eq:bar_A}
  {\mcA}(\phi)= \left[
  \frac{\mcA^{(1)}}{\phi} + \frac{\mcA^{(2)} }{1-\phi}
  + \frac{\text{d}}{\text{d}\phi} \left(\mcA^{(1)} - \mcA^{(2)}\right) \right] \,
\end{equation}
\end{subequations}
as the \emph{differential activity}. 
The derivatives of the activity $\mcA^{(i)}$ with respect to $\phi$ appear because gradients of stress enter the force balance Eqs.~\eqref{eq:force_balance_boundary1} and \eqref{eq:force_balance_boundary2}, while the dependencies on the activity $\mcA^{(i)}$ are a consequence of treating the system as a biphasic mixture, i.e.\ weighting the stress contributions by the respective volume fractions; see Supplemental Material~\cite{SM}, II for more details on Eq.~\eqref{eq:bar_A}. The specific form of the activities $\mcA^{(i)}$ depends on the system of interest.

As our first example, we consider a one dimensional, biphasic mixture composed of a (Kelvin) viscoelastic solid, (s), and a fluid phase, (f), with the constitutive equations:
\begin{subequations}\label{eq:cell_stress_FULL}
  \begin{align}\label{eq:cell_stress}
    \sigma^{\text{s},\text{p}} &= \lambda\p_x u  + \zeta \p_x v  \, ,\\
    \sigma^{\text{f},\text{p}} &= 0 \, ,
  \end{align}
\end{subequations}
where the one dimensional solid displacement and velocity  are  $u$ and $v=\text{d} u/\text{d}t\equiv \dot u$; the fluid velocity is given by $-v^{} \phi/ \left( 1-\phi \right)$.  In eq.~\eqref{eq:cell_stress}, $\lambda$ denotes the Lam\'e coefficient and $\zeta$ is the bulk  solid viscosity.  The viscous stress in the fluid phase can be approximated to zero since fluid strains are negligible relative to solid strains on length scales above the pore size, and in the systems of interest, the solid viscosity typically exceeds the fluid viscosity by several orders in magnitude~\cite{spiegelman1987simple}. Since diffusive transport of constituents in this solid-fluid mixture is expected to be slow compared to solid momentum transport, we consider the limit of small diffusivities and use rescalings of length and time scales not containing the diffusion constant. Specifically, we rescale time and length  as $t \to (\zeta/\lambda) \, t$ and $x \to \ell \, x$ with
$\ell=\sqrt{\zeta/\Gamma_0}$, so that velocities $v \to v \, \ell \lambda/\zeta$ and the scaled equations read:
\begin{subequations}\label{eq_model_one_full_equations}
\begin{align}
  \label{eq_Eq_first}
  \p_t \phi &= -\p_x (\phi \dot u ) + \tilde D \p_x^2 \phi \, ,\\
  \label{eq_Eq_second}
  0&= \p_x \left( \phi \p_x u + \phi \p_x v \right)
  + \phi \tilde A(\phi) \,  \p_x \phi - \frac{\phi}{1-\phi} \dot u \, .
\end{align}
\end{subequations}
There are two dimensionless parameters in Eqs.~\eqref{eq_model_one_full_equations}, measuring the strength of  differential activity, $\tilde A = {\mcA}/\lambda$ and  diffusivity, $\tilde D =D \Gamma_0/\lambda$.

\begin{figure*}[tb]
  \centering
  \includegraphics[width=1.0\textwidth]{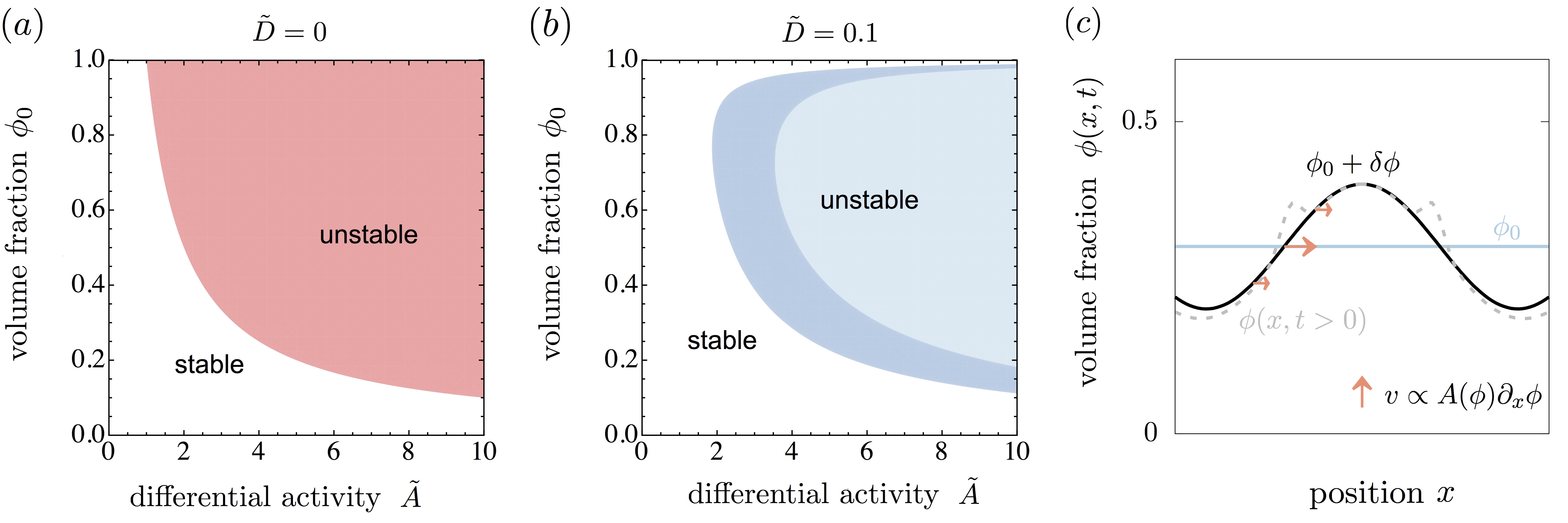}
  \caption{\label{fig:Fig_2} (a,b) Stability diagrams as a function of volume fraction
  $\phi$ and non-dimensional differential activity $\tilde A$ obtained from the linear stability analysis of Eqs.~\eqref{eq:all_equations}. 
  The diagram indicate the
  parameter regimes where the homogeneous base state is stable or unstable.
   Diagrams with constant $\tilde A$ and for diffusivities (a) $\tilde
  D=0$ and (b) $\tilde D=0.1$. The colored regions depict $\Re(\omega_+)>0$.
  Red indicates an instability where $\Im(\omega_k)=0$, while for dark/light
  blue corresponds to $\Im(\omega_k)\not= 0$ for all $q$/or for a finite
  waveband. 
  (c) Illustration of the instability mechanism driven by
  differential activity $\mcA(\phi)$. Small perturbations in volume fraction, $\phi_0
  \to \phi_0+\delta \phi$, are amplified since differential activity causes a
  drift velocity that points toward the maximum of a local inhomogeneity of
  volume fraction. To lowest order in $q$, this velocity (red horizontal arrows) scales as $ v \propto {\mcA} \p_x \phi$ (Eq.~\eqref{eq_Eq_second}), further increasing the initial perturbation. Because the velocity decreases to zero close to the extrema of the perturbation ($\p_x \phi=0$), spike may emergence (gray dashed line, $\phi(x,t>0)$), which can move inward by diffusion and thereby further amplify the instability. 
  }
\end{figure*}

To understand the stability of a homogenous base state given by $u=u_0$, $\dot u =0$ and  $\phi=\phi_0$, we perturb the volume fraction of the phases and the displacement and expand the perturbation in terms of Fourier modes of the form $\propto \text{e}^{\omega t + iqx}$ and linearize the equations above. Calculating the largest growing mode to linear order in diffusivity $\tilde D$
and to the fourth order in the  wavenumber $q$ gives
\begin{align}
  \label{eq:Re_largest_mode_expanded}
  \Re(\omega_+)&\simeq \frac{q^2}{B} \left[ -1 +\tilde A \phi_0 \left( 1 -  \frac{ \tilde D \, B }{\tilde A \phi_0 -1}\right) \right] \\
   & -\frac{q^4}{B^2} \left(\tilde A \phi_0 -1 \right) \, , \nonumber
\end{align}
for $\phi_0\in[0,1]$ and $\tilde A \phi_0 >1$, and where $B=1/(1-\phi_0)$. We see that there is a long wave length instability with $\Re[\omega_+(q)]>0$ leading to growth of the homogeneous state. The instability is driven by differential activity $\mcA$ which competes with frictional momentum transfer between the phases, and diffusion of displacement, velocity and volume fraction.  At the onset of the instability where spatial inhomogeneities in strain are negligible, 
long wavelength perturbations in volume fraction are amplified because differential activity causes a solid drift velocity $\dot u$ that points toward the maximum of a local inhomogeneity of volume fraction (Fig.~\ref{fig:Fig_2}(c)). This drift scales as $ \dot u \propto \tilde \mcA \p_x \phi$ to lowest order in $q$ (Eq.~\eqref{eq_Eq_second}. If $\mcA>0$, the velocity is parallel to the gradient in solid fraction and thus leads a local increase in the solid volume fraction. The velocity at onset of the instability is zero at the local maximum of the inhomogeneity ($\p_x \phi=0$). This causes the emergence of spikes in the volume fraction around the initial inhomogeneity where $\p_x \phi$ is largest. These spikes can move inward due to diffusion and amplify the initial perturbation (see movies in Supplemental Material~\cite{SM},V).

\begin{figure*}[tb]
  \centering
  \includegraphics[width=1.0\textwidth]{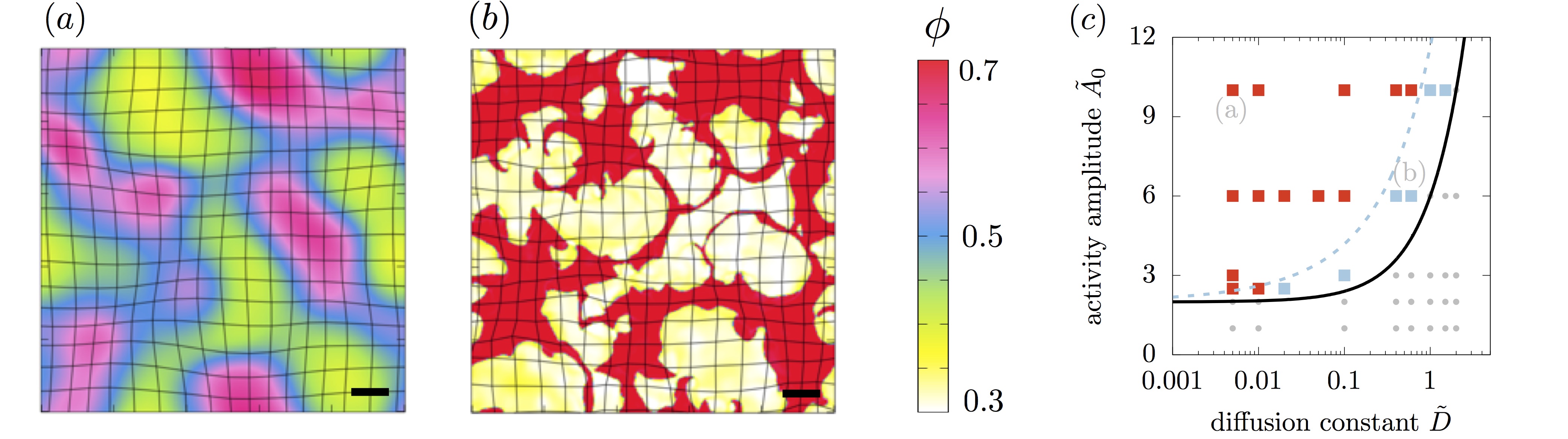}
  \caption{\label{fig:Fig_3} Patterns and phase diagram in two dimensions
  obtained from numerically solving Eqs.~\eqref{eq:all_equations} (see Supplemental Material~\cite{SM}, IV,V for details).
  (a,b) Two representative snapshots of patterns observed in our active
  poroelastic model. Black bar depicts the unit length
  $\ell=\sqrt{\zeta/\Gamma_0}$. Parameters:  $\phi_0=0.5$, and (a)
  $\tilde D=0.4$ and $\tilde A_0=6$, (b) $\tilde D=0.005$, $\tilde A_0=10$.
The black lines depict the displacements of the solid phase.
   (c) Phase diagram as a
  function of non-dimensional diffusivity $\tilde D$ and activity amplitude
  $A_0$.  Squares indicate parameters where the numerical solution
  shows the emergence a spatial-temporal pattern. Blue/red squares correspond
  to pattern morphologies reminiscent of (a)/(b). The black lines depict the
  result from the linear stability analysis for Poisson ratio $\nu \approx 0.5$.
  Parameters above the black line are linearly unstable. The parameters between
  the black and blue dashed lines correspond to oscillatory modes for all
  unstable wavenumbers.
   In the numerics we considered a specific choice of the activity function (Supplemental Material~\cite{SM}, II) to confine the range of volume fraction $\phi$ (color bar). This choice ensures the approximate validity of linear elasticity. }
\end{figure*}

When the diffusivity vanishes, i.e. $\tilde D=0$, the instability occurs for $\tilde A> \tilde A_c$ with $\tilde A_c = 1/\phi_0$ denoting the critical activity (in real units: $A_c = \lambda/\phi_0$). It is asymmetric with respect to volume fraction and the instability vanishes for $\phi_0 \to 0$ (Fig.~\ref{fig:Fig_2}(a)). The origin
of this asymmetry arises from the difference in passive properties of the two phases (Eqs.~\eqref{eq:cell_stress_FULL}). Symmetry in volume fraction can for
example be restored if both phases are treated as fluids, or as  viscoelastic material with equal transport coefficients. The growth rate of the largest
growing mode, $\omega_+(q)$, is real for all wavenumbers if $\tilde D=0$ which indicates a non-oscillatory growth of modes (see Supplemental
Material~\cite{SM}, III, for plots of $\omega_k(q)$).

For non-zero diffusivity, the critical activity increases (Eq.~\eqref{eq:Re_largest_mode_expanded} and Fig.~\eqref{fig:Fig_3}(c) black line). The term $B=1/(1-\phi_0)$ connected to viscous transport causes the instability to vanish also at large volume fraction (Fig.~\ref{fig:Fig_2}(b)). In addition, for $\tilde D \not= 0$, the growth rate $\omega_k(q)$ can have a non-zero imaginary part.  At the transition boundary between the stable and unstable regions, the growth rate is complex for all wavenumbers (dark blue/gray in Fig.~\ref{fig:Fig_2}(b)). However, deep in the unstable regime, the growth rate
becomes real for small $q$ but there remains a complex and unstable band of wavenumbers (light blue/gray in Fig.~\ref{fig:Fig_2}(b)). The
width of these band of wavenumbers decreases to zero as the diffusivity approaches zero (see Supplemental Material~\cite{SM}, III).

These two different characteristics in the growth rate obtained from the linear stability analysis indicate that nonlinear evolution of the patterns might also differ in these regimes. To investigate the pattern dynamics we numerically solved the non-linear equations in one and two dimensions; see Supplemental Material~\cite{SM}, II, IV,V for definitions of the used activity functions,  details on the numerics, and movies. In one dimension and the limit of zero diffusion, we find that the volume fraction and displacement steadily grow; a behavior that is consistent with a real dispersion relation. In the regime of a purely complex dispersion relation domains of high and low volume fraction exhibit a tendency to synchronously oscillate with a frequency that is roughly determined by the time to diffuse the size of a domain. On longer time-scales this oscillating state can spontaneously break the left-right symmetry and the domains collectively move in one direction reminiscent of  traveling fronts found in  fluid-fluid biphasic matter in the presence of osmotic forces~\cite{cogan2012marginal}. In contrast, in the mixed case where the dispersion relation is real and complex, the domains of high and low volume fraction separated by sharp interfaces seem to drift while they undergo fusion and break-up events. In two dimensions we observe a similar dynamics. For parameters closer to the transition line where all unstable modes are oscillatory, the system shows a pulsatory type of pattern (Fig.~\ref{fig:Fig_3}(a)). Deep in the unstable regime of the stability diagram (e.g.\ low $\tilde D$ and high activity amplitude $A_0$) domains with sharp and roughened interfaces drift, split and fuse (Fig.~\ref{fig:Fig_3}(b)). The onset of the instability and the two pattern morphologies determined numerically match the results obtained via linear stability (Fig.~\ref{fig:Fig_3}(c)). However, 
in two dimensions, we do not observed a collectively moving state.

Our  theory qualitatively reproduces  observations in superprecipitated active systems~\cite{mori2011intracellular, Bendix20083126,
Needleman_microtubule_contraction}. After the onset of the instability and for low diffusivities, the spatial standard deviation of displacement and volume fraction steadily increases and saturates once the system reaches its quasi stationary state. Concomitantly, the velocity decreases to zero.
According to our numerical results the inter-phase diffusion  destablizes segregated domains on long time-scales and causes oscillatory patterns. We suppose that diffusion in superprecipitated active gels is typically too weak to observe the oscillatory type of assembly/disassembly dynamics in the simulations.

We have shown that differential activity can serve as a mechanism defining a novel class of instabilities in physics,  complementing differential size~\cite{asakura1954interaction}, differential shape~\cite{onsager1949effects}, and differential adhesion~\cite{flory1942thermodynamics,huggins42,steinberg1970does}. The mechanism of differential activity might play an essential role in various kinds of biological systems including cell sorting processes in tissues,  disintegration and macroscopic contractions in super-precipitated systems and patterns in the cellular cortex or the cytoplasm.   Analogously, differential activity  characterizes the physical difference between two phases leading to the destablization of the homogeneous thermodynamic state. Our theory  generalizes one-component active fluid approaches (e.g.~\cite{bois2011pattern}) to two phases that can segregate due to the presence of active stress~\cite{radszuweit2013intracellular, radszuweit2014active}. Activity and the interactions between the phases can cause an instability leading to patches where solid or fluid matter is enriched, respectively.  This instability is driven by differential activity which competes with elastic, frictional and diffusive transport processes. Though we have illustrated the instability  for a specific set of constitutive  equations (Eqs.~\eqref{eq:cell_stress_FULL}) the existence of the instability is generic, i.e.\ it can occur for any combination of passive mechanical properties of the phases.

\newpage
\section*{acknowledgements}
We would like to thank Jakob L\"ober and Amala Mahadevan for stimulating discussions and Yohai Bar Sinai, David Fronk and Nicholas Derr for insightful comments and feedback on the manuscript. C.A.W.~ thanks the German Research Foundation (DFG) for financial support. This research was supported in part by the National Science Foundation under Grant No.\ NSF PHY-1125915.  C.H.R.~was supported by the Applied Mathematics Program of the U.S.~Department of Energy (DOE) Office of Advanced Scientific Computing Research under contract DE-AC02-05CH11231. L. M. was partially supported by fellowships from the MacArthur Foundation and the Radcliffe Institute. L.M. acknowledges partial financial support from NSF DMR 14-20570 and NSF DMREF 15-33985.

\clearpage
\section*{SUPPLEMENTAL MATERIAL}

\setcounter{table}{0}
\renewcommand{\thetable}{S\arabic{table}}%
\setcounter{figure}{0}
\renewcommand{\thefigure}{S\arabic{figure}}
\setcounter{equation}{0}
\renewcommand{\theequation}{S\arabic{equation}}

\title{Supplemental Material:\\ Differential-activity driven instabilities in biphasic active matter}

\author{Christoph A.\ Weber}
\affiliation{Paulson School of Engineering and Applied Sciences, Harvard
University, Cambridge, MA 02138, USA}
\author{Chris H.\ Rycroft}
\affiliation{Paulson School of Engineering and Applied Sciences, Harvard
University, Cambridge, MA 02138, USA}
\affiliation{Mathematics Group, Lawrence Berkeley National Laboratory,
Berkeley, CA 94720, USA}
\author{L.\ Mahadevan}
\affiliation{Paulson School of Engineering and Applied Sciences, Department of Physics, Department of Organismic and Evolutionary Biology, Harvard
University, Cambridge, MA 02138, USA}

\maketitle

\section{Origin of diffusion in biphasic matter}
In our model for biphasic active matter we introduced a relative diffusive flux
$j^{(1)}_\alpha= -j^{(2)}_\alpha =: j_\alpha$ in the transport equations of
volume fraction, $\p_t \phi^{(i)} = - \p_\alpha \left(\phi^{(i)}
v^{(i)}_\alpha + j^{(i)}_\alpha\right)$, though the velocity $v^{(i)}_\alpha$
already captures the movement of each phase. Below we will discuss one possible
mechanism for  how such a diffusive flux can
emerge and that it can be written as $j_\alpha=-D \p_\alpha \phi$, where $D$
denotes the diffusion constant.

This relative flux can for example stem from rare unbinding events of
components that belong to one of the phases. Here, a bound state refers to
filaments connected to other filaments by molecular motors. Molecular motors
can provide linkage and exert active forces on the filament phase. Unbinding
means that the constituents lose connection to other filaments due to unbinding
of molecular motors and can thus diffuse freely. Suppose the components inside
phase 1 can unbind with a rate $\gamma_2$, while  binding events occur with a rate
$\gamma_1$ and their occurrence is proportional to the fraction of unbound material and the
concentration of unbound molecular motors $c_\text{M}$. Unbound molecular
motors and unbound filaments diffuse with a diffusion constant $D_\text{M}$
and $D_\text{off}$. Assuming conservation of bound plus unbound filaments, the
extended transport equations of phase 1 are
\begin{align}
\label{eq:cm}
  \p_t c_\text{M} &= D_\text{M} \nabla^2 c_\text{M} - m \gamma_1 \phi_\text{off} c_\text{M} + m \gamma_2 \phi \, , \\
  \label{eq:unbound}
    \p_t \phi_\text{off} &= D_\text{off} \nabla^2 \phi_\text{off} - \gamma_1 \phi_\text{off} c_\text{M} + \gamma_2 \phi \, , \\
  \label{eq:phi_ext}
  \p_t \phi_\text{} &= \nabla \cdot ( \phi \vec v^{(1)} ) + \gamma_1 \phi_\text{off} c_\text{M} - \gamma_2 \phi \, ,
\end{align}
where $m$ is a constant factor accounting for binding of multiple motors per
filament.

We will now simplify the set of equations above in the limit of rare unbinding
events and fast diffusion of the unbound components, $D_\text{off} \gg \gamma_2
L^2$, where $L$ denotes the system size. This limit ensures that
$\phi_\text{off}(\vec x) \ll \phi(\vec x)$ for all positions $\vec x$. Let us
assume that diffusion of molecular motors relative to diffusion of unbound
filaments is fast ($D_\text{M} \gg D_\text{off}$) such that that
$c_\text{M}(\vec x)$ is constant. Then Eq.~\eqref{eq:cm} gives
\begin{equation}\label{eq:off}
  \phi_\text{off} \simeq \frac{\gamma_2 \phi}{\gamma_1 c_\text{M}} \, .
\end{equation}
If diffusion of unbound components is fast enough on the time-scale during
which the volume fraction $\phi$ changes, we can neglect the dynamics of
$\phi_\text{off}$, i.e.\ $\p_t \phi_\text{off} \simeq 0$, giving
\begin{equation}
 \gamma_1 \phi_\text{off} c_\text{M} - \gamma_2 \phi
 \simeq D_\text{off} \nabla^2 \phi_\text{off} \, .
 \end{equation}
Inserting Eq.~\eqref{eq:off} into the above equation, and substituting the
terms describing the binding/unbinding events in Eq.~\eqref{eq:phi_ext}, one
finds
\begin{equation}
  \p_t \phi_\text{} = \nabla \cdot ( \phi \vec v^{(1)} ) + D \nabla^2 \phi_\text{} \, ,
\end{equation}
where the diffusion constant $D=D_\text{off} \gamma_2/(\gamma_1 c_\text{M})$.
In summary, the diffusion of unbound components can effectively lead to a
diffusion of the bound components $\phi$ with a diffusive flux that can be
written as $j_\alpha=-D \p_\alpha \phi$.

A similar argument can be constructed for loosely connected tissues such as
mesenchymal tissues. The difference is only that the bound cells are active
themselves, and instead of external molecular motors one could introduce the
fraction of unbound cell-to-cell connections.

\section{Properties and possible choices of differential activity}
\subsection{Properties of differential activity}
Differential activity is the driver for the instability discussed in our
letter. In this section we discuss some basic properties of the differential
activity function,
\begin{equation}\label{eq:bar_A}
  {\mcA}(\phi)= \left[
  \frac{\mcA^{(1)}}{\phi} + \frac{\mcA^{(2)} }{1-\phi}
  + \frac{\text{d}}{\text{d}\phi} \left(\mcA^{(1)} - \mcA^{(2)}\right) \right] \, .
\end{equation}
Equation \eqref{eq:bar_A} implies that differential activity does not vanish
for equal activities, $\mcA^{(1)}=\mcA^{(2)}$. However, differential activity
can vanish even for non-zero activity in each phase, $\mcA^{(i)}\not= 0$. One
possibility is that activities cancel within the same phase,
i.e.~$\mcA^{(1)}=\mcA_{0,1}/\phi$ and $\mcA^{(2)}=\mcA_{0,2}/(1-\phi)$, where
$\mcA_{0,i}$ is some constant. The other possibility is that the activity of
one phase cancels the activity of the other phase, i.e.\ $\mcA^{(1)}=
\mp\mcA_{0} \phi$ and $\mcA^{(2)}= \pm\mcA_{0} (1-\phi)$, with $\mcA_{0}$
denoting some constant.

In the main text we discuss the case of constant differential activity. The
differential activity ${\mcA}$ can be constant and non-zero for $\mcA^{(1)}=
\pm \mcA_{0,1} \phi$ and $\mcA^{(2)} = \pm \mcA_{0,2} (1-\phi)$, or
$\mcA^{(1)}=\mcA^{(2)}=\mcA_0 \phi (1-\phi)$.
\newpage

\subsection{Possible choices of differential activity}
\begin{figure}[h]
  \centering
  \includegraphics[width=0.75\textwidth]{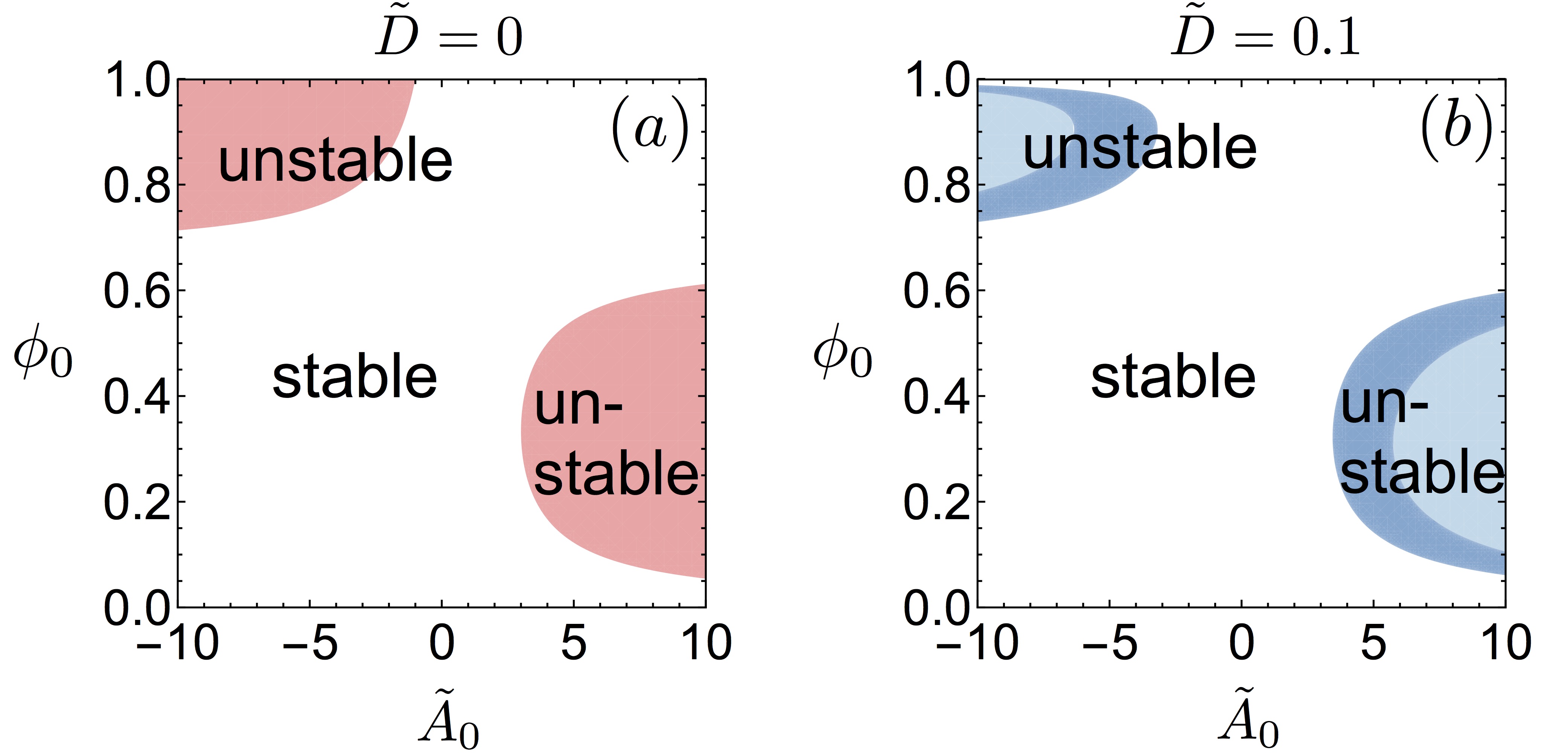}
  \caption{\label{fig:Fig_S12} Stability diagrams indicating the parameter
  regimes where the homogeneous base state is stable or unstable for the case
  of an asymmetric choice of activity, $\mcA^{(2)}=0$ and $\mcA^{(1)}=\mcA_0
  \phi_0 (1-\phi_0)$ with $\tilde{\mcA}_0 = \mcA_0/\lambda$, and (a) $\tilde
  D=0$ and (b) $\tilde D=0.1$. The colored regions depict $\Re(\omega_+)>0$.
  Dark and light blue correspond to $\Im(\omega_k)\ne 0$ for all $q$, or for a
  finite waveband, respectively.}
\end{figure}

The shape and dependencies of the activities depend on the particular system of
interest. Next to a constant differential activity one could consider an
asymmetric case where phase 2 is passive ($\mcA^{(2)}=0$) and phase 1 is active
with $\mcA^{(1)} =\mcA_0 \, \phi (1-\phi)$, leading to a differential activity
${\mcA}=\mcA_0 (2-3\phi)$ (Eq.~\eqref{eq:bar_A}). Here, $\mcA_0$ characterizes
the amplitude of the activity. In this case the generation of active stress
vanishes in the absence of phase 1 or 2, respectively. This asymmetric case is
qualitatively motivated by super-precipitated systems and
tissues~\cite{rauzi2010planar, maha_cytoplasm_poroelastic_material,
mori2011intracellular, Bendix20083126, Needleman_microtubule_contraction,
radszuweit2013intracellular, radszuweit2014active}, where one phase is passive
(intestinal fluid) while the other phase is active (filaments, cells).
Moreover, activity requires a non-zero fraction of active components
($\mcA^{(1)} \propto \phi$ ) while there could exit an inhibitory mechanism as
the active components get crowded ($\mcA^{(1)} \propto -\phi^2$). For such an
activity function linear stability suggests that an instability occurs if the
activity amplitude $\mcA_0$ is larger than the critical activity $\tilde
A_{0,c} = \left( \phi_0(2-3\phi_0) \right)^{-1}$ (for vanishing diffusivity
$\tilde D=0$). Interestingly, the instability can occur for both, positive
(expansions) and negative (contractions) activity amplitudes $\tilde A_{0}$
(Fig.~\eqref{fig:Fig_S12}). We tested this choice of activities in our
one-dimensional numerical studies and found that the emergence of
spatial-temporal patterns is consistent with the results from the linear
stability analysis. However, for parameters deeply in the unstable region of
the stability diagram, large local strain, $\p_x u$, can build up during the
pattern formation violating the small strain limit of the used Hooke's law.
Thus we considered an activity function which ensures that domains cannot
exhibit porosities below $\phi_\text{min}$ or above $\phi_\text{max}$. This
also restricts the system to small or moderate strains. The activities thus
read
\begin{align}\label{eq:activity_of_choice}
  \mcA^{(1)} &=\mcA_0 \left[ \left(\phi-\phi_0\right) - a \left(\phi-\phi_0\right)^3 \right] \, ,\\
  \mcA^{(2)} & =0 \, ,
\end{align}
where $\mcA_0$ denotes the activity amplitude (non-dimensional activity
amplitude $\tilde{\mcA}_0={A}_0/\lambda$). It sets the scale of the active
stress, but not the fixed points where the activity $\mcA^{(1)}$ vanishes to
zero. The latter is determined by the parameter $a>0$. Its value restricts the
dynamics within the volume fractions $ \phi_\text{min}= \phi_0-1/\sqrt{a}$ and
$\phi_\text{max} = \phi_0 +1/\sqrt{a}$. In our letter we fix the mean volume
fraction to $\phi_0=0.5$ and $a=20$, leading to $\phi_\text{min} \approx 0.276$
and $\phi_\text{max} \approx 0.724$.

\newpage
\cleardoublepage

%
%

\section{Growth rates as a function of wavenumber for different parameters}
\begin{figure}[h]
  \centering
  \includegraphics[width=1.0\textwidth]{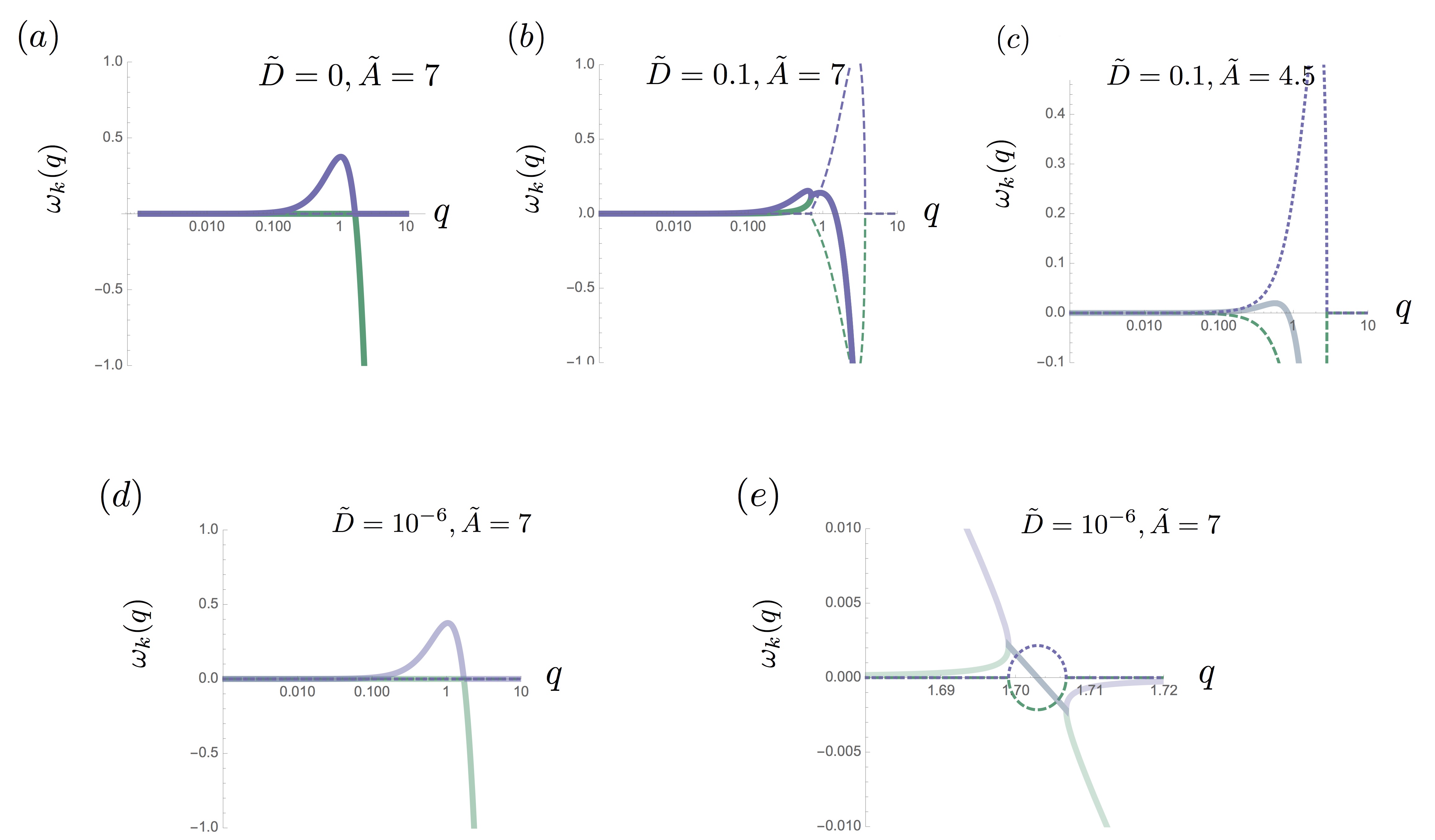}
  \caption{\label{fig:Fig_S1} Growth rate $\omega_k(q)$ as a function of
  wavenumber for different parameters for our model one-dimensional model
  discussed in the main text. Solid lines indicate the real part of the growth
  rate, $\Re[\omega_k(q)]$, while the dashed lines correspond to the imaginary
  part, $\Im[\omega_k(q)]$. Parameters are $\phi_0=0.35$ and the remaining
  parameters are shown on each plot. (a) For $\tilde D=0$, the growth rate of
  the largest growing mode, $\omega_+(q)$ (purple), is real for all wavenumbers
  indicating a non-oscillatory growth of modes. (b) As the diffusivity $\tilde
  D$ increases, the growth rate becomes real for large $q$ but there remains a
  complex growth rate for a finite waveband at larger wavenumbers (light blue
  region in the phase diagram, main text, Fig.~2b). (c) Close to the
  transition where the instability vanishes (here we have decreased $\tilde A$
  relative to (b)), the growth rate is complex for all wavenumbers (dark blue
  region in the phase diagram, main text, Fig.~2b). (d) However, deeply in
  the region of the phase diagram where the instability occurs, the width of
  the real waveband (as shown in (b)) decreases to zero as the diffusivity
  approaches zero. (e) This behavior is nicely visible in the wavenumber
  zoom-in corresponding to (d).}
\end{figure}

\newpage

\section{Details of numerical implementation}

\subsection{Dynamical equations considered in the numerical studies}

For our numerical studies we consider a biphasic mixture of a fluid (f) and
solid-like (s) phase. 
Each phase $(i)$ is described by the hydrodynamic variables: 
velocity $v^{(i)}_\alpha$, volume fraction $\phi^{(i)}$ and displacement
$u^{(i)}_\alpha$ with $\p_t u^{(i)}_\alpha = v^{(i)}_\alpha$.
For the numerical integration of the dynamical equations 
we  introduce an inertia term on the left hand side of Eqs.~(2a) and
(2b) (main text)
to make the numerical implementation straightforward (time evolution is easier than the  force balance constraint).
 However,
the quantitative influence on the solutions is expected to be negligible if the
mass density $\rho$ is small enough~\footnote{Specifically, the influence on
the numerical integration of the added inertia terms are negligible if $\hat
\rho = \rho v_0^2/\lambda \ll 1$.}. 
The full non-linear equations for the active biphasic system considered in the numerical studies read
\begin{subequations}
\begin{align}\label{eq:force_balance_boundary1}
  \p_t \phi &= - \p_\alpha \left( \phi v^\text{s}_\alpha \right) + D \p_\alpha^2 \phi \, , \\
  \label{eq:incompressibility_cond}
  0&= \p_\alpha \left[ \phi \, v^\text{s}_\alpha + \left( 1-\phi \right) v^\text{f}_\alpha \right] \, , \\
  \label{eq:dotueq}
  \p_t u_\alpha &= v^\text{s}_\alpha\, ,\\
  \rho \phi \p_t v^\text{s}_\alpha &=\p_\beta \left( \phi \sigma_{\alpha \beta}^\text{s} \right) - \phi \p_\alpha p + \Gamma_0 \phi(1-\phi) \left( v^\text{f}_\alpha - v^\text{s}_\alpha \right) \, , \\
  \label{eq:force_balance_boundary2}
  \rho (1-\phi) \p_t v^\text{f}_\alpha&=\p_\beta \left( (1-\phi) \sigma_{\alpha \beta}^\text{f} \right) - (1-\phi) \p_\alpha p - \Gamma_0 \phi(1-\phi) \left( v^\text{f}_\alpha - v^\text{s}_\alpha \right) \, ,
\end{align}
\end{subequations}
where $p$ denotes the pressure ensuring the incompressibility condition Eq.~\eqref{eq:incompressibility_cond}.
The stress tensors of the solid (s) and fluid (f) phases are
\begin{align}
\label{eq:stress_solid}
  \sigma^{\text{s}}_{\alpha \beta} = &
  \lambda \, \delta_{\alpha \beta} \p_\gamma u_\gamma
  + G \left( \p_\alpha u_\beta + \p_\beta u_\alpha \right) \\
  \nonumber
  + & \bar \lambda \, \delta_{\alpha \beta} \p_\gamma v^\text{s}_\gamma +
  \eta \left( \p_\alpha v^\text{s}_\beta + \p_\beta v^\text{s}_\alpha \right)   +
  \mcA_0 \left[ \left(\phi-\phi_0\right) - a \left(\phi-\phi_0\right)^3 \right] \delta_{\alpha \beta}\, , \\
  \sigma^{\text{f}}_{\alpha \beta} &= 0 \, ,
\end{align}
where $\lambda$ is the first Lam\'e parameter and $G=\lambda(1-2\nu)/(2\nu)$ is
the shear modulus, and $\nu$ denotes the Poisson ratio. 
Moreover, $\bar \lambda$ is
the bulk viscosity and $\eta$ is the shear viscosity. 
The second viscosity $\zeta=\bar \lambda+2\eta/3$. 
We used the choice of the activity function given in Eq.~\eqref{eq:activity_of_choice}.
For the fluid phase we
neglected contributions proportional to velocity gradients because they are
expected to be small on scales above the size of the solid pores. In addition,
the solid viscosities typically exceeds the fluid viscosities by several orders
of magnitude.

We rescale length and time as $x \to \ell \cdot x$ with
$\ell=\sqrt{\bar{\lambda}/\Gamma_0}$, and $t \to (\ell/v_0) \cdot t$, and
consider a velocity rescaling $v^{(i)}_\alpha \to v_\alpha^{(i)} \cdot v_0 $.
The dimensional pressure and stress are $ p \to p \cdot \lambda$ and
$\sigma_{\alpha \beta}^{(i)} \to \sigma_{\alpha \beta}^{(i)} \cdot \lambda$.
After this rescaling there are five dimensionless parameters in more than one
dimension, namely $ \tilde A_0 = {\mcA_0}/\lambda$,$ \tilde D =D /(\ell v_0)$,
$\tilde G =G/\lambda$,  
 ${V}= v_0 \bar \lambda /(\lambda \ell)= v_0
\sqrt{\Gamma_0 \bar \lambda}/\lambda$, and $\tilde \eta =V \eta/\bar{\lambda}$.
In one dimension the two shear parameters ($\tilde G$, $\tilde
\eta$) do not exist leading to three parameters. In the main text we considered the special case where the velocity scale $V=1$  leaving us with two
non-dimensional parameters, namely the activity amplitude $\tilde A_0$ and the
diffusivity $\tilde D$.

The dimensionless equations
used for numerical discretization are
\begin{subequations}
\begin{align}\label{eq:force_balance_boundary1n}
  \p_t \phi &= - \p_\alpha \left( \phi v^\text{s}_\alpha \right) + \tilde D \p_\alpha^2 \phi \, , \\
  \label{eq:incompressibility_condn}
  0&= \p_\alpha \left[ \phi \, v^\text{s}_\alpha + \left( 1-\phi \right) v^\text{f}_\alpha \right] \, , \\
  \label{eq:dotueqn}
  \p_t u_\alpha &= v^\text{s}_\alpha\, ,\\
  \label{eq:force_balance_boundary1_soft}
  \hat \rho \phi \p_t v^\text{s}_\alpha &=\p_\beta \left( \phi \sigma_{\alpha \beta}^\text{s} \right) - \phi \p_\alpha p + 
  \phi(1-\phi) \left( v^\text{f}_\alpha - v^\text{s}_\alpha \right) \, , \\
  \label{eq:force_balance_boundary2_soft}
  \hat \rho (1-\phi) \p_t v^\text{f}_\alpha &= - (1-\phi) \p_\alpha p - 
   V \, \phi(1-\phi) \left( v^\text{f}_\alpha - v^\text{s}_\alpha \right) \, ,
\end{align}
\end{subequations}
where the dimensionless mass density is $\hat \rho = \rho v_0^2/\lambda$. 

\subsection{Numerical methods used to solve dynamical equations in two dimensions}

In the numerical integration we consider periodic boundary conditions. 
The two-dimensional simulations are
performed with a custom C++ code and use an $N\times N$ grid for 
non-dimensional system size $(L/\ell)\times(L/\ell)$, where $\ell=\sqrt{\bar{\lambda}/\Gamma_0}$ is the unit length and $L$ is the system size.
 The spatial derivatives are discretized
using second-order finite-differences. The time derivatives are discretized by
an Euler scheme with a time increment of $\Delta t$ and a discrete time $t=n
\cdot \Delta t$, where $n$ denotes the $n$-th time step. 

In two dimensions, the essential step in the numerical integration is the calculation of the pressure in a way that the incompressibility holds (Eq.~\eqref{eq:incompressibility_condn}). To this end, we apply a variant
of Chorin's projection method as used in the integration of the incompressible Navier--Stokes
equations~\cite{chorin68}. 
This method amounts to splitting the integration of
Eqs.~\eqref{eq:force_balance_boundary1_soft} and
\eqref{eq:force_balance_boundary2_soft} into three parts. The first part
calculates intermediate velocities $v^{\text{s},*}_\alpha$ and
$v^{\text{f},*}_\alpha$ by neglecting the pressure gradient and relaxing the
incompressibility constraint. These velocities are then used in the second
part to compute the pressure via solving a Poisson problem. In the final
part, the pressure is employed to project the velocities to satisfy the
incompressibility constraint.

The first part of the
projection step can thus be written as
\begin{subequations}
\begin{align}
  \hat \rho \frac{v^{\text{s},*}_\alpha -v^{\text{s}}_\alpha (n)}{\Delta t} &=\phi(n)^{-1}\p_\beta \left( \phi(n) \sigma_{\alpha \beta}^\text{s}(n) \right) + V \, \left(1-\phi\left(n\right)\right) \left( v^\text{f}_\alpha(n) - v^\text{s}_\alpha(n) \right) \, , \\
  \hat \rho \frac{v^{\text{f},*}_\alpha -v^{\text{f}}_\alpha (n)}{\Delta t} &= - V \, \phi(n) \left( v^\text{f}_\alpha(n) - v^\text{s}_\alpha(n) \right) \, .
\end{align}
\end{subequations}
The second step can be derived by writing down the third step of the projection method,
\begin{subequations}
\begin{align}\label{eq:vnp1_s}
  \hat \rho \frac{v^{\text{s}}_\alpha(n+1) -v^{\text{s},*}_\alpha }{\Delta t}
  &= - \p_\alpha p(n+1) \, ,\\
  \label{eq:vnp1_f}
  \hat \rho \frac{v^{\text{f}}_\alpha(n+1) -v^{\text{f},*}_\alpha }{\Delta t}
  &= - \p_\alpha p(n+1) \, .
\end{align}
\end{subequations}
The third step requires the knowledge of the pressure $p(n+1)$ at time step
$n+1$ and computes the incompressible velocities $v^{\text{s}}_\alpha(n+1)$ and
$v^{\text{f}}_\alpha(n+1)$ at time step $n+1$. The pressure is computed in the
second step. Since $v^{\text{s}}_\alpha(n+1)$ and $v^{\text{f}}_\alpha(n+1)$
obey the incompressibility condition, multiplying Eq.~\eqref{eq:vnp1_s} by
$\phi(n)$ and Eq.~\eqref{eq:vnp1_f} by $(1-\phi(n))$, and adding up both
equations, applying the divergence and using the incompressibility condition
for the incompressible velocities (Eq.~\eqref{eq:incompressibility_condn}), we
find
\begin{align}\label{}
   \p_\alpha^2 p(n+1)
   & = \frac{\hat \rho}{\Delta t} \p_\alpha \left[ \phi(n) \, v^{\text{s},*}_\alpha + \left(1-\phi\left(n\right)\right) \, v^{\text{f},*}_\alpha \right] \, .
\end{align}
The equation above represents the second step of the projection step. It is an
elliptic problem, and is solved using a custom C++ implementation of the
geometric multigrid method. Solving this equation gives the pressure $p(n+1)$,
which is required in the third steps as described above.

\begin{table*}[t]
  \centering
  \begin{tabular}{|c|c|c|c|c|c|c|c|c|c|c|}
    \hline
    $\phi_0$ & $\tilde D$ & $\tilde A_0$ & $\nu$ & $V$ & $\tilde \eta$
    & $\hat \rho$ & $a$ \\
    \hline
    $0.5$ & $0.005$--$5$ & $0$--$10$ & $0.48$ ($\tilde G \approx 0.0417$)
    & 0.06 & $0.02$ 
    & $0.02$ & $20$ \\
    \hline
  \end{tabular}
  \caption{A list of fixed parameters used to numerically solve
  Eq.~\eqref{eq:final_equs} in two spatial dimensions with the solid stress
  tensor given in Eq.~\eqref{eq:stress_solid}. For the numerical results please
  refer to main text, Fig.~3 and movies
  Sec.~\ref{sect:videos}.\label{tab:values_parameters}}
\end{table*}

In summary, given the fields $\phi(n)$, $v^{\text{s}}_\alpha (n)$,
$v^{\text{f}}_\alpha (n)$ and $u_\alpha(n)$ at time step $n$, the full
projection based algorithm in time is
\begin{subequations}\label{eq:final_equs}
\begin{align}
  \hat \rho \frac{v^{\text{s},*}_\alpha -v^{\text{s}}_\alpha (n)}{\Delta t} &=\phi(n)^{-1}\p_\beta \left( \phi(n) \sigma_{\alpha \beta}^\text{s}(n) \right) + V \, \left(1-\phi\left(n\right)\right) \left( v^\text{f}_\alpha(n) - v^\text{s}_\alpha(n) \right) \, , \\
  \hat \rho \frac{v^{\text{f},*}_\alpha -v^{\text{f}}_\alpha (n)}{\Delta t} &= - V \, \phi(n) \left( v^\text{f}_\alpha(n) - v^\text{s}_\alpha(n) \right) \, , \\
  \p_\alpha^2 p(n+1)
  & = \frac{\hat \rho}{\Delta t} \p_\alpha \left[ \phi(n) \, v^{\text{s},*}_\alpha + \left(1-\phi\left(n\right)\right) \, v^{\text{f},*}_\alpha \right] \, ,\\
  \hat \rho \frac{v^{\text{s}}_\alpha(n+1) -v^{\text{s},*}_\alpha }{\Delta t}
  &= - \p_\alpha p(n+1) \, ,\\
  \hat \rho \frac{v^{\text{f}}_\alpha(n+1) -v^{\text{f},*}_\alpha }{\Delta t}
  &= - \p_\alpha p(n+1) \, , \\
  \frac{\phi(n+1)-\phi(n)}{\Delta t} &= - \p_\alpha \left( \phi(n) v^\text{s}_\alpha(n) \right) + \tilde D \p_\alpha^2 \phi(n) \, , \\
  \frac{u_\alpha(n+1)-u_\alpha(n)}{\Delta t} &= v^\text{s}_\alpha(n)\, .
\end{align}
\end{subequations}
The spatial derivatives are implemented using a second-order finite difference
discretization (not shown). Parameters for the integration in two spatial
dimensions are given in Table~\ref{tab:values_parameters}.

\subsection{Numerical solution of dynamical equations in one dimensions}

In one dimension, the projection steps are not necessary since the
incompressibility condition (Eq.~\eqref{eq:incompressibility_condn}) implies a
linear relationship between the velocities. With appropriate boundary
conditions, $v^{f}= -v^{s} \phi/ \left( 1-\phi \right)$. Therefore, the
pressure is determined and can be substituted. For the one-dimensional studies
we considered a velocity scale $V=1$ for simplicity, leading to
\begin{subequations}\label{eq:model_one_full_equations}
\begin{align}
  \label{eq:Eq_first}
  \p_t \phi &= -\p_x (\phi v ) + \tilde D \p_x^2 \phi \, ,\\
  \label{eq:Eq_second}
  \p_t u &= v\, ,\\
  \label{eq:Eq_third}
  \hat \rho \frac{\phi}{1-\phi} \p_t v&= \p_x \left( \phi \p_x u \right)
  + \p_x \left( \phi \p_x v \right)
  + \phi ( \mcA/\lambda) \,  \p_x \phi - \frac{\phi}{1-\phi} v^\text{s} \, ,
\end{align}
\end{subequations}
where we abbreviated $v=v^s$ and $\mcA(\phi)$ denotes the differential activity
(main text, Eq.~(3c)). We verified our one-dimensional results by our
implementation as outlined above with the results obtained from a spectral
based solver (XMDS2~\cite{dennis2013xmds2}) and a finite element solver (Comsol
Multiphysics\textsuperscript{\textregistered} software~\cite{comsol}).
The movies in 1D were rendered using the Comsol software.


\section{Movie description}\label{sect:videos}

We attached two movies for the one-dimensional (1D)
equations~\eqref{eq:model_one_full_equations}, and two for the two-dimensional
(2D) equations~\eqref{eq:final_equs}. If not stated below, parameters 
are given in Table~\ref{tab:values_parameters}.

\begin{itemize}
  \item[(1)] 1D: oscillating dynamics and collective motion. $\tilde D=0.4$, $\phi_0=0.5$, $\tilde A_0=6$, duration $T_\text{}=500$, $L/\ell=50$, grid points $N=500$.
  \item[(2)] 1D: drifting domains undergoing fusion and break-up. $\tilde D=0.005$, $\phi_0=0.5$, $\tilde A_0=10$, duration $T_\text{}=250$, $L/\ell=50$, grid points $N=500$.
  \item[(3)] 2D: pulsatory-type of dynamics.  $\tilde D=0.4$, $\phi_0=0.5$, $\tilde A_0=6$, duration $T_\text{}=100$, $L/\ell=10$, grid points $N=256$.
  \item[(4)] 2D:  drifting domains undergoing fusion and break-up.
   $\tilde D=0.005$, $\phi_0=0.5$, $\tilde A_0=10$, duration $T_\text{}=100$, $L/\ell=10$, grid points $N=256$.
\end{itemize}


\begin{thebibliography}{10}
\bibitem{footnote2}
The osmotic pressure can also be absorbed into a renormalized activity, which
  only shifts the transition to larger activities.
\end{thebibliography}

\end{document}